\documentclass[prb,twocolumn,superscriptaddress,showpacs,
amsmath,amssymb]{revtex4}

\usepackage{graphicx} 
\usepackage{dcolumn} 
\usepackage{hyperref}

\def\bG{{\bf G}}

\def\bj{{\bf j}}
\def\bk{{\bf k}}

\def\bq{{\bf q}}

\def\bG{{\bf G}}
\def\bQ{{\bf Q}}
\def\bS{{\bf S}}

\def\b0{{\bf 0}}
\def\bGam{{\bf\Gamma}}
\def\bSg{{\bf\Sigma}}

\def\bra{\langle}
\def\ket{\rangle}
\def\up{\uparrow}
\def\down{\downarrow}

\def\eps{\epsilon}

\def\Gam{\Gamma}

\def\Lam{\Lambda}

\def\sg{\sigma}
\def\Sg{\Sigma}

\def\phib{\bar\phi}
\def\psib{\bar\psi}

\def\sgn{{\rm sgn}}

\begin{document}

\title{Superconductivity in the two-dimensional $t$-$t'$-Hubbard model}

\author{Andreas Eberlein}
\affiliation{Max Planck Institute for Solid State Research,
 D-70569 Stuttgart, Germany}
\author{Walter Metzner}
\affiliation{Max Planck Institute for Solid State Research,
 D-70569 Stuttgart, Germany}

\date{\today}

\begin{abstract}
Using a recently developed renormalization group method for
fermionic superfluids, we determine conditions for $d$-wave 
superconductivity in the ground state of the two-dimensional 
Hubbard model at moderate interaction strength, and we compute 
the pairing gap in the superconducting regime.
A pairing instability signaled by a divergent flow in the Cooper 
channel leads to a superconducting state in all studied cases.
The next-to-nearest neighbor hopping $t'$ plays a crucial role in 
the competition between antiferromagnetism and superconductivity.
A sizable $t'$ is necessary to obtain a sizable pairing gap.
\end{abstract}

\maketitle


\section{Introduction}

Shortly after the discovery of high-temperature superconductivity
in layered cuprate compounds, Anderson \cite{anderson87} suggested
that the two-dimensional Hubbard model contains the essence of the 
electron dynamics in the copper-oxygen planes.
While it may not describe all relevant aspects of the system,
the Hubbard model definitely captures its most prominent property,
that is, $d$-wave superconductivity in the vicinity of 
antiferromagnetic order.\cite{scalapino95} 
Convincing evidence for superconductivity in the Hubbard model at 
weak and moderate coupling strengths has been established by 
self-consistent or renormalized perturbation expansions, 
\cite{bickers89,neumayr03,kyung03,raghu10} and from functional
renormalization group flows.\cite{zanchi97,halboth00,honerkamp01}
At stronger coupling, embedded cluster methods \cite{maier05} 
yield superconducting states in a large density range, if magnetic
order is excluded,\cite{maier00,khatami08,gull13} and otherwise
surprisingly extended regions of superconductivity with a sizable
pairing gap coexisting with antiferromagnetism.
\cite{lichtenstein00,capone06,aichhorn06}
Variational Monte Carlo calculations with superconducting trial
wave functions revealed a substantial energy gain from $d$-wave
pairing in a wide density range in the strong coupling regime.
\cite{yokoyama13}
On the other hand, unbiased quantum Monte Carlo (QMC) simulations
frequently detected enhanced superconducting fluctuations, but 
only rarely evidence for long-range order.\cite{qmc}

At weak and moderate coupling the functional renormalization
group (fRG) is probably the most powerful method for studying
the interplay of magnetism and superconductivity in 
two-dimensional lattice electron models.\cite{metzner12}
In this method, approximations are derived by truncating an
exact flow equation for the effective action, where the flow
parameter $\Lam$ is usually an energy scale controlling the
successive integration of fluctuations.\cite{frg}
The fRG treats all fluctuation contributions to the effective 
two-particle interaction and self-energy on equal footing and 
in the thermodynamic limit. 
The $d$-wave pairing instability generated by magnetic 
fluctuations in the two-dimensional Hubbard model emerges
already within the lowest order (one-loop) truncation.
\cite{zanchi97,halboth00,honerkamp01}
The instability is signaled by a divergence of the effective
two-particle interaction in the Cooper channel at a critical
cutoff scale $\Lam_c$.

Antiferromagnetic fluctuations are the main mechanism for 
$d$-wave pairing interactions, at least for a moderate Hubbard 
interaction, but magnetism also competes with superconductivity, 
since magnetic order (static or fluctuating) leads to gaps in
the electronic spectrum.
From the early fRG flows \cite{zanchi97,halboth00,honerkamp01} 
the competition between antiferromagnetism and superconductivity 
could not be decided unambiguously in a sizable density range 
where both channels develop large effective interactions, since 
the flow had to be stopped at the scale at which the effective 
interaction diverges, and it was not clear whether the leading
divergence is a reliable indicator for the prevailing type of 
order.

To continue the flow beyond the critical scale one has to allow
for spontaneous symmetry breaking.
One possibility is to introduce a bosonic order parameter field
by a Hubbard-Stratonovich decoupling of the interaction.
This approach to symmetry breaking in the fRG has already been 
applied to antiferromagnetic \cite{baier04} and superconducting 
\cite{strack08,friederich10} states in the Hubbard model.
The choice of a specific decoupling procedure of the Hubbard 
interaction introduces a certain bias, which leads to ambiguities 
in cases with competing instabilities.
Alternatively, one may work with a purely fermionic flow,
which is the route we take here.
In the fermionic fRG, a relatively simple one-loop truncation 
with self-energy feedback \cite{katanin04} solves mean-field 
models of symmetry breaking such as the reduced BCS model exactly, 
although the effective interaction diverges at $\Lam_c$.
\cite{salmhofer04}
For the {\em attractive} Hubbard model, this truncation yields 
results for the pairing gap in good agreement with earlier 
estimates at weak and moderate coupling strength.\cite{gersch08}
Recently, an improved parametrization of the interaction 
vertex in a fermionic superfluid, which fully exploits spin
rotation invariance and parametrizes singularities by a 
single momentum and frequency variable, was derived.
\cite{eberlein10,eberlein13}
It is based on an extension of a decomposition of the 
normal-state vertex in charge, magnetic, and pairing channels 
\cite{karrasch08,husemann09} to the superfluid state.
This new parametrization was also applied to the attractive
Hubbard model, and a comprehensive understanding of the 
behavior of the flowing effective interaction was obtained.
\cite{eberlein13}

In the present work, we use the fermionic fRG to detect and 
analyze superconductivity in the ground state of the 
two-dimensional {\em repulsive}\ Hubbard model.
We find that a diverging $d$-wave pairing interaction always 
leads to a superconducting state, and we compute the $d$-wave 
gap as a function of doping for various choices of the 
next-to-nearest neighbor hopping $t'$.
The results reveal the crucial role of $t'$ in the competition 
between magnetism and superconductivity.

The paper is structured as follows. In Sec.~II we describe the
fRG equations for an unbiased detection and analysis of $d$-wave 
superconducting states in the two-dimensional Hubbard model. 
In Sec.~III we present results for effective interactions, 
critical scales and the ground state pairing gap. 
A short summary and final remarks in Sec.~IV close the presentation.


\section{Model and Method}

In standard second-quantization notation the Hubbard model
\cite{montorsi92} is described by the Hamiltonian
\begin{equation}
 H = \sum_{\bj,\bj',\sg} t_{\bj\bj'} 
 c_{\bj\sg}^{\dag} c_{\bj'\sg} +
 U \sum_{\bj} n_{\bj\up} n_{\bj\down} \; ,
\end{equation}
where $\bj,\bj'$ label lattice sites and $\sg$ is the spin orientation.
For nearest and next-to-nearest neighbor hopping on a square 
lattice with amplitudes $-t$ and $-t'$, respectively, the
Fourier transform of the hopping matrix yields a dispersion
$\eps_{\bk} = -2t(\cos k_x + \cos k_y) - 4t' \cos k_x \cos k_y$.
We set $t=1$, which defines our unit of energy.

The partition function and generating functionals for correlation 
functions can be written as functional integrals over anticommuting
fields $\psi_{k\sg}$ and $\psib_{k\sg}$, where $k = (k_0,\bk)$ 
comprises Matsubara frequencies and momenta.
The generating functional $\Gam$ for one-particle irreducible 
vertex functions, also known as {\em effective action}, is given by 
the Legendre transform of the generating functional for connected
Green functions.\cite{negele87}
Adding a suitable regulator term to the quadratic part of the
bare action, one can define a scale dependent effective action
$\Gam^{\Lam}$ that interpolates smoothly between the bare action
$\cal S$ at the highest scale $\Lam_0$ and the final effective
action $\Gam$ for $\Lam \to 0$.
The flow of $\Gam^{\Lam}$ obeys an exact functional flow equation,
\cite{wetterich93} from which one can derive a hierarchy of
flow equations for the vertex functions.

To describe a superfluid state, it is convenient to use a
representation in terms of {\em Nambu}\/ fields $\phi_{ks}$, 
$\phib_{ks}$ defined as
$\phi_{k+} = \psi_{k\up}$, $\phib_{k+} = \psib_{k\up}$, 
$\phi_{k-} = \psib_{-k\down}$, $\phib_{k-} = \psi_{-k\down}$.
To quartic order in the fields, the scale dependent effective 
action for a spin-singlet superfluid has the general form
\cite{fn1}
\begin{eqnarray}
 \Gam^{\Lam}[\phi,\phib] &=& \Gam^{(0)\Lam} -
 \sum_k \sum_{s_1,s_2} \Gam_{s_1s_2}^{(2)\Lam}(k) \,
 \phib_{k s_1} \phi_{k s_2} \nonumber \\
 &+& \frac{1}{4}  \sum_{k_1,\dots,k_4} \sum_{s_1,\dots,s_4} 
 \Gam_{s_1s_2s_3s_4}^{(4)\Lam}(k_1,k_2,k_3,k_4) 
 \nonumber \\ 
 &\times&
 \phib_{k_1s_1} \phib_{k_2s_2} \phi_{k_3s_3} \phi_{k_4s_4}
 \; .
\end{eqnarray}
For systems with (unbroken) spin-rotation invariance, only terms
with an equal number of $\phi$ and $\phib$ fields contribute.
The Nambu vertex $\Gam_{s_1s_2s_3s_4}^{(4)\Lam}(k_1,k_2,k_3,k_4)$ is
nonzero only for $k_1 + k_2 = k_3 + k_4$.
The Nambu components of the 2-point function 
$\Gam_{s_1s_2}^{(2)\Lam}(k)$ form a $2\times2$ matrix 
$\bGam^{(2)\Lam}(k)$.
Its matrix inverse is the Nambu propagator
\begin{equation} \label{bG}
 \bG^{\Lam}(k) = 
 \left( \begin{array}{cc}
 G^{\Lam}(k) & F^{\Lam}(k) \\
 F^{*\Lam}(k) & -G^{\Lam}(-k)
 \end{array} \right) ,
\end{equation}
where $G^{\Lam}(k) = - \bra \psi_{k\sg} \psib_{k\sg} \ket$
and $F^{\Lam}(k) = - \bra \psi_{k\up} \psi_{-k\down} \ket$.
The Dyson equation 
$(\bG^{\Lam})^{-1} = (\bG_0^{\Lam})^{-1} - \bSg^{\Lam}$
relates the full propagator $\bG^{\Lam}$ to the 
self-energy $\bSg^{\Lam}$ and the bare regularized 
propagator $\bG_0^{\Lam}$ given by
\begin{equation} \label{G0}
 \left[ \bG_0^{\Lam}(k) \right]^{-1} = \left(  
 \begin{array}{cc}
 ik_0 - \xi_{\bk} + R^{\Lam}(k_0) &
 \Delta_0(k) \\
 \Delta_0^*(k) &
 ik_0 + \xi_{\bk} + R^{\Lam}(k_0)
 \end{array} \right) ,
\end{equation}
where $\xi_{\bk} = \eps_{\bk} - \mu$, and $\Delta_0(k)$ is a 
small initial gap added to the bare action to trigger the 
symmetry breaking.
It can be chosen small enough to avoid a discernible effect 
on the gap at the end of the flow.
The regulator function
$R^{\Lam}(k_0) = i \, \sgn(k_0) \sqrt{k_0^2 + \Lam^2} - ik_0$
replaces frequencies $k_0$ with $|k_0| \ll \Lam$ by 
$\sgn(k_0) \Lam$ and thus confines the bare propagator to a
size of order $\Lam^{-1}$.
The self-energy matrix has the form
\begin{equation} \label{bSg}
 \bSg^{\Lam}(k) = 
 \left( \begin{array}{cc}
 \Sg^{\Lam}(k) & \Delta_0(k) - \Delta^{\Lam}(k) \\
 \Delta_0^*(k) - \Delta^{\Lam*}(k) & - \Sg^{\Lam}(-k)
 \end{array} \right) ,
\end{equation}
where $\Delta^{\Lam}(k)$ is the flowing gap function.

The Nambu self-energy obeys the exact flow equation
\begin{equation} \label{floweq_Sg}
 \frac{d}{d\Lam} \Sg_{s_1s_2}^{\Lam}(k) =
 \sum_{k'} \sum_{s'_1,s'_2} S_{s'_2s'_1}^{\Lam}(k')
 \Gam^{(4)\Lam}_{s_1s'_1s'_2s_2}(k,k',k',k) \; ,
\end{equation}
where
$\bS^{\Lam}(k) = \left. \frac{d}{d\Lam} 
 \bG^{\Lam}(k) \right|_{\bSg^{\Lam} \, {\rm fixed}}$.
The flow of the Nambu vertex $\bGam^{(4)\Lam}$ is approximated
by a one-loop truncation with self-energy feedback 
\cite{katanin04} where contributions from three-particle
interactions leading to two- and higher loop terms are
neglected. This approximation is exact for mean-field models
such as the reduced BCS model.\cite{salmhofer04}
The flow equation for $\bGam^{(4)\Lam}$ is then given by a sum 
of three one-loop diagrams corresponding to the particle-particle, 
direct and crossed particle-hole channel, respectively.
\cite{eberlein10,eberlein13}

The parametrization of the Nambu vertex is based on an
extension of the channel decomposition devised initially
for the normal state \cite{karrasch08,husemann09} to a
spin-singlet superfluid. 
The fluctuation contributions to the normal effective 
interaction are decomposed in a charge, a magnetic, and a 
pairing contribution, where possible singular momentum
and frequency dependences of the corresponding coupling
functions $C_{kk'}^{\Lam}(q)$, $M_{kk'}^{\Lam}(q)$, and 
$P_{kk'}^{\Lam}(q)$ are isolated in the variable $q$, 
which is either a momentum transfer or a conserved total 
momentum (or frequency).
In a superfluid state also anomalous interactions appear.
A coupling function $W_{kk'}^{\Lam}(q)$ describes the 
destruction or creation of four electrons, while another 
function $X_{kk'}^{\Lam}(q)$ captures anomalous processes 
with three ingoing electrons and one outgoing electron, 
or vice versa.\cite{eberlein10,eberlein13}

We adopt a static approximation for the vertex, that is,
we discard the frequency dependences of the coupling
functions.
The $q_0$-frequency dependence of the coupling functions is 
crucial for capturing the dynamics of infrared singularities 
associated with the Goldstone boson,\cite{eberlein13} but 
this has little impact on the gap function.
By fixing the phase of the gap at zero, the gap function and
all (static) coupling functions are real.
In the normal and anomalous pairing channels it is convenient
to use amplitude and phase coupling functions defined as
$A_{\bk\bk'}^{\Lam}(\bq) = 
 P_{\bk\bk'}^{\Lam}(\bq) + W_{\bk\bk'}^{\Lam}(\bq)$ and
$\Phi_{\bk\bk'}^{\Lam}(\bq) = 
 P_{\bk\bk'}^{\Lam}(\bq) - W_{\bk\bk'}^{\Lam}(\bq)$,
respectively.
The dependence of the coupling functions on the fermionic 
momenta $\bk$ and $\bk'$ is parametrized by an expansion 
in the simplest $s$-wave and $d$-wave form factors, 
$s_{\bk} = 1$ and $d_{\bk} = \cos k_x - \cos k_y$,
respectively:
\begin{eqnarray}
 C_{\bk\bk'}^{\Lam}(\bq) &=& 
 C_s^{\Lam}(\bq) + 
 C_d^{\Lam}(\bq) d_{\bk} d_{\bk'} \; ,
 \nonumber \\
 M_{\bk\bk'}^{\Lam}(\bq) &=& 
 M_s^{\Lam}(\bq) + 
 M_d^{\Lam}(\bq) d_{\bk} d_{\bk'} \; ,
 \nonumber \\
 A_{\bk\bk'}^{\Lam}(\bq) &=& 
 A_s^{\Lam}(\bq) + 
 A_d^{\Lam}(\bq) d_{\bk} d_{\bk'} \; ,
 \nonumber \\
 \Phi_{\bk\bk'}^{\Lam}(\bq) &=& 
 \Phi_s^{\Lam}(\bq) + 
 \Phi_d^{\Lam}(\bq) d_{\bk} d_{\bk'} \; ,
 \nonumber \\
 X_{\bk\bk'}^{\Lam}(\bq) &=& 
 X_{sd}^{\Lam}(\bq) d_{\bk'} + 
 X_{ds}^{\Lam}(\bq) d_{\bk} \;.
\end{eqnarray}
For the first four coupling functions, mixed $s$-$d$-terms are
very small and can be neglected.\cite{fn_hs} On the other
hand, the last one is dominated by mixed terms, while 
diagonal $s$-$s$- and $d$-$d$-terms are negligible here. 
The neglected terms are fully absent in a mean-field model
with reduced $s$- and $d$-wave interactions in the forward
scattering and pairing channels.\cite{eberlein13a}
The $\bq$-dependences of the coupling functions cannot be 
parametrized accurately by simple functions and are therefore
discretized on a two-dimensional grid.

Inserted into the flow equation (\ref{floweq_Sg}), a static
real vertex entails a frequency-independent real self-energy.
The momentum dependence of its normal component is weak and
has no important effects.\cite{giering12} 
We therefore approximate $\Sg^{\Lam}$ by a constant. 
For the momentum dependence of the gap function we use the 
simplest $d$-wave ansatz
$\Delta^{\Lam}(\bk) = \Delta^{\Lam} d_{\bk}$, and 
correspondingly $\Delta_0(\bk) = \Delta_0 \, d_{\bk}$.

The flow of the coupling functions, self-energy and gap is
obtained by projecting the right hand sides of the flow
equations on the ansatz via Fermi surface averages. 
\cite{eberlein13}
Deviations from the Ward identity relating gap and vertex
are eliminated during the flow by another projection.
\cite{eberlein13,eberlein13a}


\section{Results}

We now present results based on a numerical solution of the
flow equations.
In Fig.~1 we show the flow of various coupling functions at
fixed momenta for a moderate interaction strength $U=3$, 
a next-to-nearest neighbor hopping $t'=-0.25$, and density 
$n=0.9$. For these parameters the ground state is a $d$-wave 
superconductor with a gap amplitude $\Delta(0,\pi) = 2\Delta^{\Lam=0} = 0.047$.
\begin{figure}[t]
\begin{center}
\includegraphics[width=7.5cm]{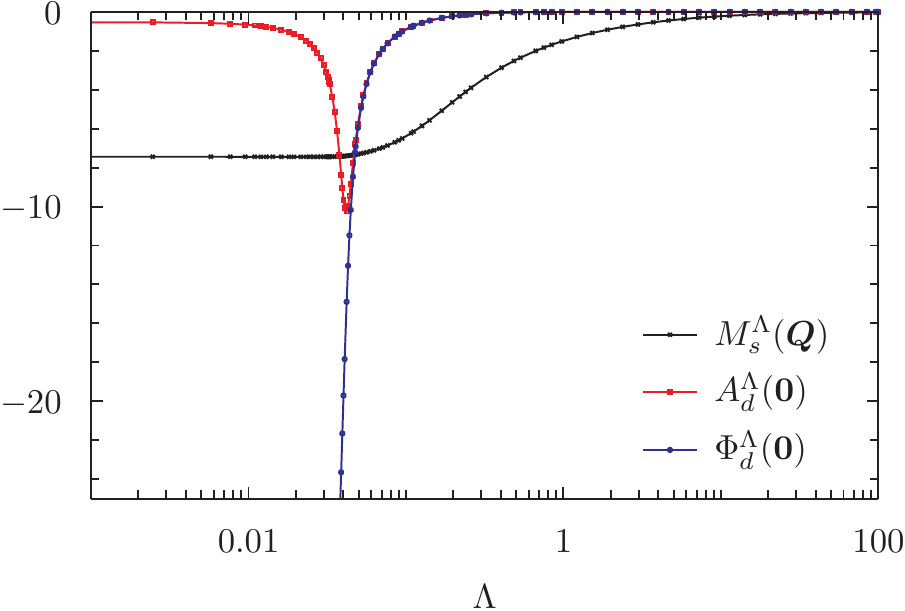} \\[5mm]
\includegraphics[width=7.5cm]{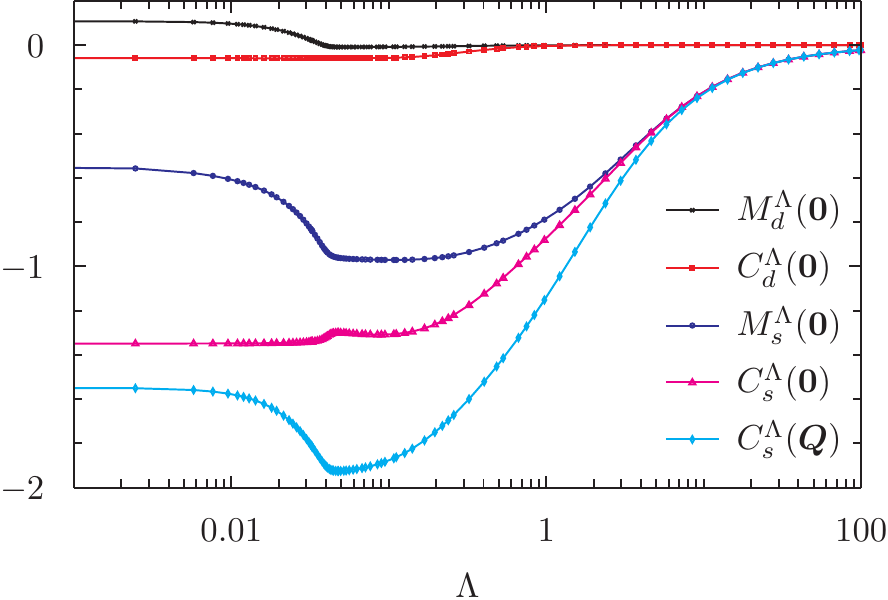}
\caption{(Color online) Flows of coupling functions for
 $U = 3$, $t' = -0.25$, and density $n = 0.9$.
 Top: Dominant magnetic and $d$-wave pairing coupling functions
 at $\bq = \bQ = (\pi,\pi)$ and $\bq = \b0$, respectively.
 Bottom: Charge coupling functions at $\bq = \b0$ and 
 $\bq = (\pi,\pi)$, and magnetic coupling functions at 
 $\bq = \b0$. Note the distinct scales on the vertical axes 
 of the top and bottom panel.}
\end{center}
\end{figure}
The pairing instability at $\Lam_c = 0.040$ is generated
mostly by antiferromagnetic fluctuations. 
The latter grow gradually already at scales well above 
$\Lam_c$, as can be seen from the flow of $M_s^{\Lam}(\bq)$
at $\bq = (\pi,\pi)$.
The $d$-wave pairing amplitude coupling $A_d^{\Lam}(\b0)$ 
exhibits a pronounced peak at the critical scale $\Lam_c$.
The presence of a small external pairing gap ($\Delta_0 = 
1.6 \times 10^{-4}$) prevents a divergence of the peak. 
The phase coupling $\Phi_d^{\Lam}(\b0)$ increases rapidly at 
$\Lam_c$ and saturates at a large final value proportional
to $\Delta_0^{-1}$.

Other coupling functions remain relatively small. 
Some examples are shown in the lower panel of Fig.~1.
In particular, the $d$-wave charge coupling function 
$C_d^{\Lam}(\bq)$ is only weakly attractive for all wave
vectors $\bq$.
A large negative $C_d^{\Lam}(\b0)$ would indicate an 
incipient $d$-wave Pomeranchuk instability \cite{halboth00,
yamase00} toward nematic order.\cite{fradkin10}
A strongly attractive $C_d^{\Lam}(\bq)$ at $\bq \neq \b0$
would signal a modulated nematic instability,\cite{holder12} 
which can also be viewed as a $d$-wave bond order.
Such an instability was shown to accompany $d$-wave pairing
near an antiferromagnetic quantum critical point.
\cite{metlitski10}
However, a recent fRG study of the Hubbard model above the 
critical scale $\Lam_c$ did not reveal any proximity to 
$d$-wave charge order,\cite{husemann12} in agreement with our results.

The leading instabilities are generically either 
antiferromagnetism or $d$-wave pairing.
In the upper panel of Fig.~2 we show the critical scale 
$\Lam_c$ as a function of ``doping'' $x = 1-n$ at a fixed
interaction strength $U=3$ for various choices of $t'$.
The doping range covers a broad regime from moderate electron 
doping to fairly large hole doping. 
Distinct symbols for $d$-wave superconductivity, commensurate 
and incommensurate antiferromagnetism indicate which coupling 
function diverges at $\Lam_c$. 
Incommensurate antiferromagnetism is signaled by a 
divergence of $M_s(\bq)$ at wave vectors of the form 
$(\pi \pm \delta,\pi)$ and $(\pi,\pi \pm \delta)$.
\begin{figure}[t]
\begin{center}
\includegraphics[width=\linewidth]{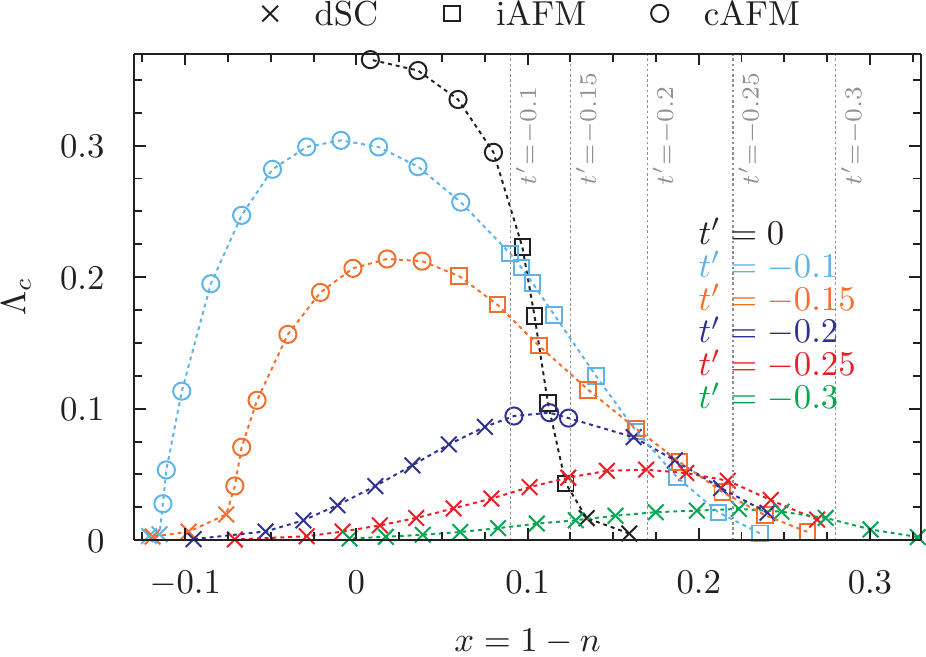} \\[5mm]
\includegraphics[width=\linewidth]{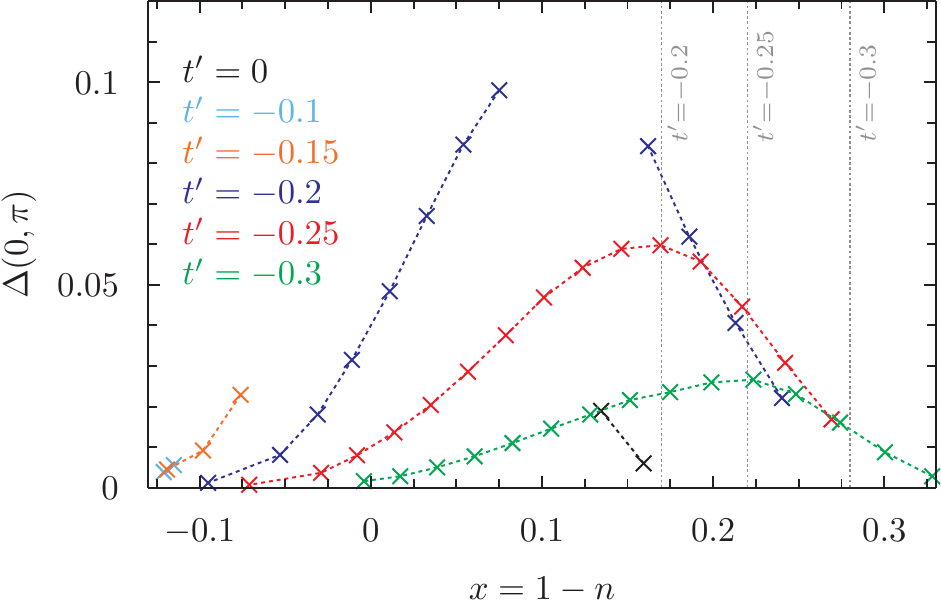} \\[5mm]
\caption{(Color) Critical scales for the leading
 instability (top) and $d$-wave gap amplitude (bottom) as a
 function of doping for $U = 3$ and various choices of $t'$.
 The leading instability is specified by different symbols
 for $\Lam_c$, and gaps are shown only in the superconducting
 regime where the flow could be continued to $\Lam = 0$.
 The dotted gray vertical lines indicate Van Hove filling 
 for different values of $t'$.}
\end{center}
\end{figure}
Note that $\Lam_c$ is maximal {\em above}\ Van Hove filling 
for all $t' < 0$.
This is due to a mutual reinforcement of different channels
in the presence of antiferromagnetic hot spots.\cite{fn2}

Whenever pairing is the leading instability, we continue
the flow to $\Lam = 0$ and compute the $d$-wave pairing gap. 
In the lower panel of Fig.~2 the resulting gap amplitudes 
$\Delta(0,\pi) = 2 \Delta^{\Lam=0}$ are plotted as a 
function of doping.
One can see that $\Delta(0,\pi)$ is comparable to $\Lam_c$.
\cite{fn3}
Fluctuations below $\Lam_c$ have little influence on the
size of the gap.
An important observation is that in all cases of a pairing 
instability at $\Lam_c$, the flow could be continued to a
superconducting state at $\Lam = 0$. 
Hence, a divergence of the vertex in the pairing channel
at $\Lam_c$ is a reliable indicator for a superconducting
state.
Previously, the leading instability was often determined 
at a scale $\Lam_* > \Lam_c$ at which the vertex exceeds 
a certain large finite value.\cite{metzner12}
This was partially motivated by concerns about the validity 
of the one-loop truncation in the regime of large effective
interactions.
However, such a supposedly cautious procedure can lead to
incorrect conclusions, since a divergence in the pairing
channel is often preceded by a regime of dominant magnetic
interactions at scales $\Lam > \Lam_c$.
On the other hand, we cannot exclude the possibility that
the superconducting state obtained from the fRG flow is
only metastable. In particular, at and near half-filling
there might be an antiferromagnetic ground state that is not 
signaled by a divergent interaction in the flow, analogously 
to a first order phase transition which is not signaled by
a divergent susceptibility.

A divergence of the magnetic coupling function at a scale
$\Lam$ below the critical scale for pairing $\Lam_c$ would
indicate magnetic order coexisting with superconductivity 
(as the leading instability).
We have never encountered such a divergence, in agreement
with a previous study based on a combination of fRG and 
mean-field theory, where cases of coexistence with a dominance 
of pairing turned out to be extremely rare.\cite{reiss07}
Vice versa, a dominant magnetic instability naturally allows for 
pairing with a smaller energy scale, when the magnetic order
does not fully gap the Fermi surface.

A superconducting state at half-filling, as obtained for 
$t'=-0.2$, is possible only for weak or moderate interactions. 
\cite{hassan08,sentef11,gull13}
At strong coupling, the half-filled system is a Mott insulator
and magnetic order is the only option for symmetry breaking.

The maximal size of the pairing gap (at ``optimal doping'') 
depends strongly on $t'$.
For $|t'| \leq 0.15$, the leading instability near half-filling
is always antiferromagnetic, and $d$-wave pairing is leading only
in a density range away from half-filling where the critical
scale and the pairing gaps are already quite small.
For $t'=0$, there is pairing with a small but visible gap around 
$x=0.15$, and, due to the particle-hole symmetry for $t'=0$, also
at $x=-0.15$ (not shown).
For $t'=-0.1$ and $-0.15$, in the pairing regime at large hole
doping, $\Lam_c$ and $\Delta$ are smaller than the resolution in 
Fig.~2, and are therefore not plotted.
The extended regime of incommensurate antiferromagnetism on
the hole doped side is due to Fermi surface nesting.
For $|t'| \geq 0.25$, $d$-wave pairing is the only instability 
for all densities in the plotted range.
The antiferromagnetism found in the hole-doping range around 
$x=0.1$ for $t'=-0.2$ is almost degenerate with superconductivity.
In this regime, we find commensurate antiferromagnetism due to 
umklapp scattering between antiferromagnetic hot spots.
The largest pairing gap is obtained for $t' = -0.2$ near that
antiferromagnetic regime for moderate hole-doping above Van Hove 
filling.
Hence, a substantial but not too large negative value of $t'$ 
is optimal for obtaining superconductivity with a large gap
in the hole-doped system.
In the weak and moderate interaction regime, where the one-loop 
truncation is a controlled approximation, the optimal size of 
$|t'|$ increases monotonically with $U$. 
Hence, we expect an optimal value $t'_{\rm opt} < -0.2$
for interactions $U > 3$.


\section{Conclusion}

We have used a fermionic fRG, with a channel decomposition that
treats charge, magnetic, and pairing interactions on equal footing,
to determine the energy scale and the nature of the leading 
instabilities in the two-dimensional repulsive Hubbard model at
a moderate interaction strength.
Depending on the model parameters, one finds divergent interactions
indicating commensurate or incommensurate antiferromagnetism, or 
$d$-wave superconductivity, as in previous fRG studies.\cite{metzner12}
A recent extension of the fRG for superfluid states allowed us
to compute the pairing gap in the superconducting regime.
A pairing instability signaled by a divergence in the Cooper
channel leads to a superconducting state in all studied cases.
We have scanned a wide parameter range, with densities ranging
from moderate electron-doping to large hole-doping,
and several choices of a next-to-nearest neighbor hopping $t'$.

The strong $t'$-dependence resulting from our fRG study is 
consistent with unbiased QMC simulations of the Hubbard model, 
where pairing turned out to be too weak to be detected at 
$t'=0$,\cite{aimi07} while evidence for superconductivity 
was found at $t'=-0.2$.\cite{yanagisawa10}
Band structure calculations by Pavarini et al.\ 
\cite{pavarini01} revealed long ago that a substantial
hopping amplitude beyond nearest neighbors is beneficial for 
high-temperature superconductivity in cuprates. 
Comparing many cuprate compounds, they found empirically that 
$T_c$ at optimal doping increases systematically with the hopping
range.


\vspace{5mm}
\begin{acknowledgments}
We would like to thank O.~K.~Andersen, K.-U.~Giering, O.~Gunnarsson, 
T.~Holder, C.~Husemann, B.~Obert, M.~Salmhofer, H.~Yamase, and
R.~Zeyher for valuable discussions.
Support from the DFG research group FOR 723 is also gratefully
acknowledged.
\end{acknowledgments}



\begin{thebibliography}{99}
\bibitem{fn1} The general ansatz for the effective action and
 the ensuing flow equations are described in detail in 
 Ref.~\onlinecite{eberlein13}.
\bibitem{fn_hs} In the normal metallic regime above $\Lam_c$ 
 this was checked in Ref.~\onlinecite{husemann09}.
\bibitem{fn2} Hot spots are intersections of the Fermi 
surface with the antiferromagnetic Brillouin zone boundary.
\bibitem{fn3} The ratio between gap and $\Lam_c$ depends 
to some extent on the form of the regulator function 
$R^{\Lam}$. For the regulator chosen here, one obtains
$\Delta/\Lam_c = 1$ in a reduced BCS model with $s$-wave
pairing.\cite{eberlein13} For a sharp frequency cutoff
one would expect smaller values of $\Lam_c$.
\end{thebibliography}
\end{document}